\documentclass{article}

\usepackage{arxiv}

\usepackage{amsmath,amssymb}
\usepackage{graphicx}
\usepackage{booktabs}
\usepackage{url}
\usepackage{hyperref}
\usepackage{algorithm}
\usepackage{algorithmic}
\usepackage{pifont}
\usepackage{makecell}
\usepackage{subcaption}
\usepackage{xcolor}

\newcommand{\bd}{\boldsymbol{d}}
\newcommand{\xmark}{\textcolor{red}{\ding{55}}}
\newcommand{\cmark}{\textcolor{green!60!black}{\ding{51}}}

\newcommand{\shorttitle}{Fast and Feasible: Permutation-based Constrained Reranking}

\title{Fast and Feasible: Permutation-based Constrained Reranking for Revenue Maximization}

\author{
  Svetlana Shirokovskikh \\
  Avito\\
  Moscow, Russia\\
  \And
  Anastasiia Soboleva\thanks{Both authors contributed equally to this research.} \\
  Avito \& MSU, AI Center and IAI \\
  Moscow, Russia \\
  \And
  Ekaterina Solodneva\footnotemark[1] \\
  MSU, AI Center \& Avito \\
  Moscow, Russia \\
  \And
  Aleksandr Katrutsa\thanks{Corresponding author.} \\
  Avito \& MSU, AI Center\\
  Moscow, Russia \\
  \texttt{amkatrutsa@gmail.com}
  \And
  Roman Loginov \\
  Avito \\
  Moscow, Russia \\
  \And
  Egor Samosvat \\
  Avito\\
  Moscow, Russia\\
}

\begin{document}

\maketitle

\begin{abstract}
Search and recommender systems have produced highly relevant search results.
A natural next step in the development of such systems in e-commerce is to rerank these results to increase the platform's revenue from paid promotion products.
However, maximizing revenue alone may degrade the user experience by reducing relevance or increasing fraud risk.
To avoid this, we state the reranking problem as an integer linear program ($ILP$) that maximizes revenue subject to per-query constraints on other metrics, e.g., relevance. 
Since solving $ILP$ exactly for every query is slow for deployment to the online service, we propose a lightweight permutation-based reranking approximation algorithm \textbf{PermR}.
At each step, the algorithm selects a pair of neighboring items and swaps them to either improve the objective or repair a violated constraint.
We evaluate \textbf{PermR} across multiple categories of a large classified platform in offline and online settings.
\textbf{PermR} achieves about 63\% of the ILP revenue improvement, within production latency limits, preserving all constraints.
In a 14-day online A/B test over 56 million search queries, \textbf{PermR} increased revenue by $2$\%.
\end{abstract}

\keywords{reranking, constrained optimization, e-commerce}

\begin{figure}[h]
\centering
  \includegraphics[width=0.9\textwidth]{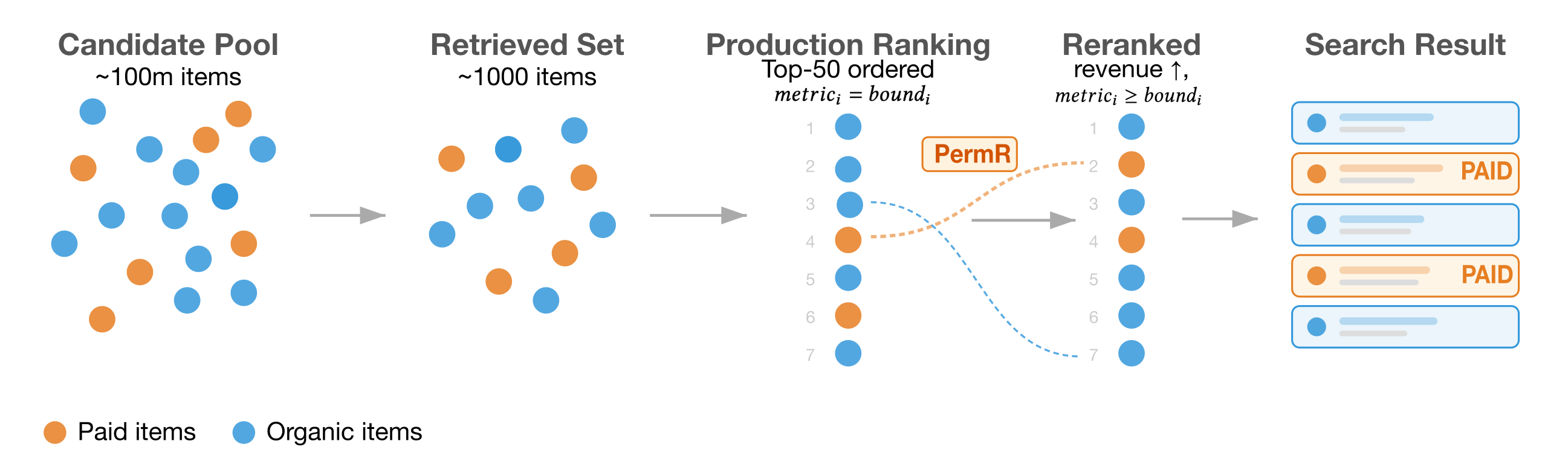}
  \caption{Multi-stage reranking pipeline with \textbf{PermR} layer on the top.}
  \label{fig::teaser}
\end{figure}

\section{Introduction}

Modern search systems typically employ a multi-stage ranking pipeline.
First, a set of candidates is retrieved, then one or more ranking stages reorder them using increasingly complex models~\cite{guo2022semantic, huang2025comprehensive}. 
One way to optimize for additional objectives is to add a reranking layer on top of the pipeline~\cite{liu2022neural}. 
Revenue optimization through reranking can increase platform revenue in e-commerce~\cite{solodneva2025rare, meng2025generative}.  
However, optimizing for revenue tends to degrade metrics such as relevance and fraud safety.
If reranking is performed by a separate model, it may depend on features that vary over time.
Thus, this model requires regular retraining whenever the features change. 

Multi-objective optimization in ranking~\cite{fatima2025multi, zheng2022survey} can be approached via scalarization~\cite{van2016balancing} or Pareto-based approaches~\cite{ribeiro2015multiobjective}.
However, scalarization is sensitive to weights, and Pareto-based methods produce multiple solutions.
In contrast, the constrained optimization approach optimizes a single target metric while bounding others~\cite{lu2024power, basu2016constrained}.  
This makes constrained optimization a natural formulation with a single solution~\cite{singh2018fairness, hao2021re}.
This approach leads to an integer linear programming (ILP) problem. 
In~\cite{lu2024power}, constraint bounds for ILP formulation are set relative to an ideal ranking, while we define constraints \emph{relative to the current production ranking}.

An ILP solution can be obtained by MOSEK~\cite{mosek} and HiGHS~\cite{huangfu2018parallelizing}, based on branch-and-bound methods~\cite{wolsey2020integer}.
However, they are slow for real-time production environments.
LP relaxation methods~\cite{wolsey2020integer} reduce solving time, while producing fractional solutions, and require careful rounding.
Study~\cite{lu2024power} proposes a faster algorithm based on the bisection method for a single constraint.
However, its extension to multiple constraints requires efficient approximate methods that find near-optimal solutions under strict latency constraints.

We consider a large-scale classifieds platform where the ranking pipeline produces high-quality search results. 
We propose a permutation-based algorithm that reranks these results as a lightweight post-processing layer (see Figure~\ref{fig::teaser}).
The algorithm iteratively swaps neighboring items to increase revenue while satisfying constraints.
Since the existing ranking is already strong, constraints are defined as maximum allowed degradation relative to the current result rather than as absolute targets.
Different constraints can consider different prefix sizes of the result list.
Table~\ref{table:comparison} summarizes the key features of our method compared to existing approaches.

\begin{table}[ht]
  \caption{Comparison of revenue-aware reranking approaches. 
  Revenue-aware models include~\cite{bae2025ranking, meng2025generative, zhuang2018globally, solodneva2025rare}. ``No training'' means the method does not require a trained ranking model. ``Multiple constraints'' means the method can simultaneously enforce bounds on several metrics. ``Time Complexity`` means the method requires production latency constraints.}
  \label{table:comparison}
  \centering
  \small
  \begin{tabular}{lccc}
    \toprule
    \textbf{Approach} & \textbf{No training} & \makecell{\textbf{Multiple}\\\textbf{constraints}} & \makecell{\textbf{Time}\\\textbf{Complexity}} \\
    \midrule
    Revenue-aware models & 
      \xmark & 
      \xmark &
      \cmark\\
    \addlinespace
    LP solver & 
      \cmark & 
      \cmark &
      \xmark \\
    \addlinespace
    LP-based algorithm~\cite{lu2024power} & 
      \cmark & 
      \xmark &
      \cmark\\
    \addlinespace
    \textbf{PermR} (Ours) & 
      \cmark & 
      \cmark &
      \cmark\\
    \bottomrule
  \end{tabular}
\end{table}

The main contributions of our work are the following
\begin{enumerate}
    \item We state the revenue-aware reranking task as an ILP problem
    \item We propose the permutation-based approximation algorithm, \textbf{PermR}, that increases revenue and satisfies constraints
    \item We compare ILP solvers and \textbf{PermR} on the industrial data 
    \item Online A/B-tests confirm the performance of \textbf{PermR} 
\end{enumerate}

\section{Problem Formulation}
The environment can be described as follows: a sequence of $T$ incoming queries arrives online, indexed by $t = 1, \dots, T$. 
For each query $q_t$, the search system returns the \textbf{SERP} $(1,\dots,N_t)$, following the notation \cite{roy2022users,sheng2025progressive} for the Search Engine Result Page.
The platform decides which item, indexed $i = 1, \dots, N_t$, and in which position, indexed by $j = 1, \dots, N_t$, to display as a response to query $q_t$.
A number of metrics $F_m,\ m = 0, 1, \dots, M$ are used to evaluate the quality of the SERP (e.g., relevance, fraud detection, metric responsive for low reputation, etc.).
Each item $i$ is characterized by its estimated contribution to each metric~$F_m$ by $f^{(i)}_m,\ m =0,1, \dots M$. 
Since we consider multiple slots, the order in which items are presented significantly affects their visibility.
We model the position bias using the Position-Based Model~\cite{chuklin2022click}, which assigns examination probabilities $\gamma_j \in [0,1]$ to each ranking position~$j$. 
We assume that~$\gamma_l$ are known to the platform and decrease with increasing $l$, i.e., $1 = \gamma_1 \geq \gamma_2 \geq \ldots \geq \gamma_{L_t}$.
So, if item $i$ is displayed in position $j$, then its estimated contribution to metric~$F_m$ is given by: $\gamma_j\cdot f^{(i)}_m$.
The goal of the platform is to maximize the primary metric $F_0$, subject to constraints on other metrics $F_m,\ m =1,\dots,M$. 
Let us focus on a particular search query $q_t$, and suppress the subscript $t$ to simplify notation.
Let $\bd$ be a binary matrix $N \times N$ such that $d_{ij} = 1$ if item $i$ is displayed in position $j$.
Then the SERP optimization problem can be formulated as an integer linear programming problem~(\ref{eq::RerankingLP}):
\begin{equation}
\label{eq::RerankingLP}
\tag{\textit{ILP}}
\begin{split}
    & \max_{\bd}  \sum_{i=1}^N\sum_{j=1}^N \gamma_j\cdot d_{ij}\cdot f_0^{(i)}\\
    \text{s.t. } 
    & \sum_{i=1}^N\sum_{j=1}^N \gamma_j\cdot d_{ij}\cdot f_m^{(i)}\geq F_m^*,\ \forall m=1,\dots,M \\
    & \sum_{i=1}^N\sum_{j=1}^K \gamma_j\cdot d_{ij}\cdot f_m^{(i)}\geq F_m^*@K,\ \forall m=1,\dots,M\\
    & \sum_{j=1}^N d_{ij} = 1,\ \forall i=1,\dots,N\\
    & \sum_{i=1}^N d_{ij} = 1,\ \forall j =1,\dots,N\\
    & d_{ij}\in \{0,1\},\ \forall i=1,\dots,N\ \forall j=1,\dots,N.\\  
\end{split}
\end{equation} 
For the bounds $F_{m}^{*}$, $m=1,\dots,M$, we use the values from a current production ranking model.
This ensures that production ranking is feasible.
The constraints on $\bd$ include:
(1) each slot $j$ can be allocated to only one item: $\sum_{i=1}^N d_{ij} = 1$, (2) each item $i$ can be displayed in only one slot $\sum_{j=1}^N d_{ij} = 1$. 
$F_m@K$ adds a constraint for top-K.

\section{PermR: Permutation-based Algorithm}
The exact solution of~\ref{eq::RerankingLP} is too slow for deployment in the real-time system.
We propose a permutation-based PermR algorithm that provides an approximate solution to~\ref{eq::RerankingLP} that satisfies constraints, achieves near-optimal objective values, and ensures low latency.

PermR performs $I$ iterations and modifies the top-$N_t$ items in the search result.
At each step, it checks if all constraints $F_m \geq F^*_m$ are satisfied.
If all constraints are satisfied, PermR attempts to improve the objective $F_0$.
Let $s_j$ be the item at position $j$ for a current permutation.
Then, PermR samples a pair of neighboring items $j, j+1$ with probability proportional to  $\max(0, f_0^{s_{j+1}} - f_0^{s_j})$.
Since $\gamma$ from~\ref{eq::RerankingLP} monotonically decreases, swapping a higher-value item into a higher position increases the objective.
If any of the constraints is violated, then PermR randomly selects from the violated constraints $F_m$.
Then it samples a neighboring pair with a probability proportional to $\max(0, f_m^{s_{j+1}} - f_m^{s_{j}})$ and applies the transposition. 

Some constraints are defined on a prefix size $K_m \leq N_t$.
In this case, there may be no transposition to improve the violated constraint $F_m$. 
Then PermR selects an item with the highest~$f_m$ and places it in the first position. 
If such an item is already in the first position, then the second-highest will be placed in the second position, and so on.
For the complete pseudo-code of PermR see Algorithm~\ref{alg::perm}.

\begin{algorithm}[!ht]
\caption{PermR}
\begin{algorithmic}[1]
\REQUIRE items $s_1,\dots, s_{N_t}$; number of iterations $I$; bounds $F^*_m$, $K_m$ size of elements for constraints on the prefix
\STATE $\pi_0\gets (s_1,\dots, s_{N_t})$, $\pi^* \gets \pi_0$
\FOR{$i=1,\dots, I$}
\STATE $V\gets \{m: F_m(\pi_{i-1}) < F^*_m\}$ \COMMENT{set of violated constraints}
\IF{$V = \emptyset$}
\STATE If $F_0(\pi_{i-1}) > F_0(\pi^*)$: $\pi^*\gets \pi_{i-1}$
\STATE $w_j\gets \max(0,\; f_0^{s_{j+1}} - f_0^{s_{j}}),\quad j=1,\dots, N_t-1$
\STATE Sample $j\sim w_j$
\STATE $\pi_i \gets (s_1, \dots, s_{j+1}, s_j, \dots, s_{N_t})$
\ELSE 
\STATE Sample $m\sim \mathcal{U}(V)$ \COMMENT{select random element from $V$}
\STATE $w_j\gets \max(0,\; f_{m}^{s_{j+1}} - f_m^{s_{j}}),\quad j=1,\dots, N_t-1$
\IF{$\exists\, j:\; w_j > 0$}
\STATE Sample $j \sim Categorical\left(\frac{w_1}{\sum_iw_i}, \dots, \frac{w_{N_{t-1}}}{\sum_iw_i}\right)$
\STATE $\pi_i \gets (s_1, \dots, s_{j+1}, s_j, \dots, s_{N_t})$
\ELSE
\STATE Let $p$ be the largest index such that $s_1,\dots,s_p$ are sorted by $f_m$ in descending order
\IF{$p \leq K_m$}
\STATE $k \gets \arg\max_{l > p} f_m^{s_l}$
\STATE $\pi_i\gets \text{Insert } s_k \text{ at position } p+1$
\ELSE
\STATE $\pi_i\gets\pi_{i-1}$
\ENDIF
\ENDIF
\ENDIF
\ENDFOR
\RETURN $\pi^*$
\end{algorithmic}
\label{alg::perm}
\end{algorithm}

\section{Experiments}
This section numerically evaluates PermR against baselines in the offline experiments and online A/B tests.
For reproducibility, we share the repository\footnote{\url{https://github.com/avito-tech/PermutationReranking}}. 
The repository includes the code and a subset of the product data used to validate the findings from Table~\ref{tab::time_ilp}.

\subsection{Production Environment}
\label{subsection:prod}
We evaluated \textbf{PermR} in a large classified platform environment.
Our goal metric $F_0$ is the aggregated expected revenue of an item (e.g., pay-per-click/per-contact) given the impression.
There are a total of 7 constraints in our production problem:
\begin{itemize}
    \item Two relevance metrics $F_1, F_2$ (e.g., $F_1$ is the predicted relevance by the ML model)
    \item Four platform's quality metrics $F_3, F_4, F_5, F_6$ (e.g., $F_3$ is fraud detection metric, $F_4$ responses to low reputation, $F_5$ is detection metric of private sellers, and $F_6$ is responsible for prepaid promotion products)
    \item An additional one constrained $F_1@5$ on the predicted relevance for the top-5.
\end{itemize}

The decay factor for position bias, $\gamma_j = 0.97^j$, is consistent with the real production environment.
Since it decays with the position, the top-ranked items contribute more to the metric values.
Therefore, the algorithm places high-revenue items up, while still preserving other constraints.
Experiments were conducted in the Goods category, which comprises approximately 200 million items.

\subsection{Offline Experiments: PermR vs ILP solvers}
In offline experiments, we solve problem~\ref{eq::RerankingLP} for the production environment in real logged data.
We aim to determine the number of iterations for PermR (Algorithm~\ref{alg::perm}) and identify the most efficient ILP solver based on selective SERPs with $N=50$ items.
Table~\ref{tab::time_ilp} presents the evaluation results of ILP solvers HiGHS~\cite{huangfu2018parallelizing}, MOSEK~\cite{mosek}, the genetic algorithm (GA) from the pymoo framework~\cite{pymoo} under different inference time limits $T$, and our PermR with varying numbers of iterations $I$.
For MOSEK, we explore the impact of thread scaling from $1$ to $12$ threads on performance.
For GA, the population size $P$ was chosen within a grid-search over \{10, 25, 50, 100\} for each time limit.
Based on Table~\ref{tab::time_ilp}, we run MOSEK with $12$ threads in further experiments. 
Note, scaling from 12 to 96 threads does not yield acceleration.
ILP solvers provide exact solutions to the problem, achieving identical revenues  (uplift of $\Delta\ \text{Rev}\ (\%) = +6.9\%$, see Table~\ref{tab::time_ilp}). 
An increase in PermR iterations from $100$ to $1000$ increases revenue by $+1.3\%$ to $+6.0\%$.
However, due to production limits on inference time ($0.05$ seconds), we set the PermR iterations to $750$, and GA under $0.05$ seconds time limit.
PermR achieves a $62\%$ approximation of the solver's revenue uplift with 750 iterations. 
On platforms with inference times up to $0.25$ seconds, PermR with 2500 iterations provides $80\%$ approximation.

\begin{table}[!h]
\centering
\caption{For queries with $N = 50$ items, we identify \#~iterations ($I$) for PermR, the time limit ($T$) for GA, and the best ILP solver from HiGHS and MOSEK~{(M)} on different threads.
ILP solvers provide an upper bound on revenue uplift ($\Delta \text{Rev} (\%) = +6.9$).
To solve ILPs, we use MOSEK with 12 threads, as it achieves the lowest average runtime.
Based on the production time limit, we set $I = 750$ and the GA time limit to $0.05$.
}
\begin{tabular}{c|cccccccc}
\toprule
 \textbf{ILPs}          &HiGHS   &M1   &M2  &M4  &M6  &M8  &M10&\textbf{M12}\\
 \textbf{Mean} time ($sec$)            &3.257 &1.963 &1.194&1.101&0.721&0.665&0.680&\textbf{0.612}\\
 $\Delta$ Rev (\%) & +6.9 & +6.9 & +6.9 & +6.9 & +6.9 & +6.9 & +6.9 & +6.9 \\
\midrule
 \textbf{PermR}        &$I=$100   &250  &500  &\textbf{750} &1000&2500&5000&10000\\
 \textbf{Max} time ($sec$)             &0.007 &0.018  &0.022 &\textbf{0.038}&0.053&0.111 & 0.231& 0.463\\

  $\Delta$ Rev (\%)    &+1.3  &+2.7   &+3.8  &+4.3   &+4.6&+5.5 & +5.7 & +6.0 \\
  \midrule
 \textbf{GA}        &$T=$0.01   &0.02  &\textbf{0.05}  &0.1 &0.25&0.5&0.75&1.0\\
  $\Delta$ Rev (\%)    &+0.10  &+0.14   &+0.48  &+0.97   &+1.80 & +2.58 & +3.10 & +3.37 \\
  \bottomrule
\end{tabular}
\label{tab::time_ilp}
\end{table}

\begin{figure}[!h]
     \centering
     \begin{subfigure}[t]{0.48\linewidth}
         \centering
         \includegraphics[width=\linewidth]{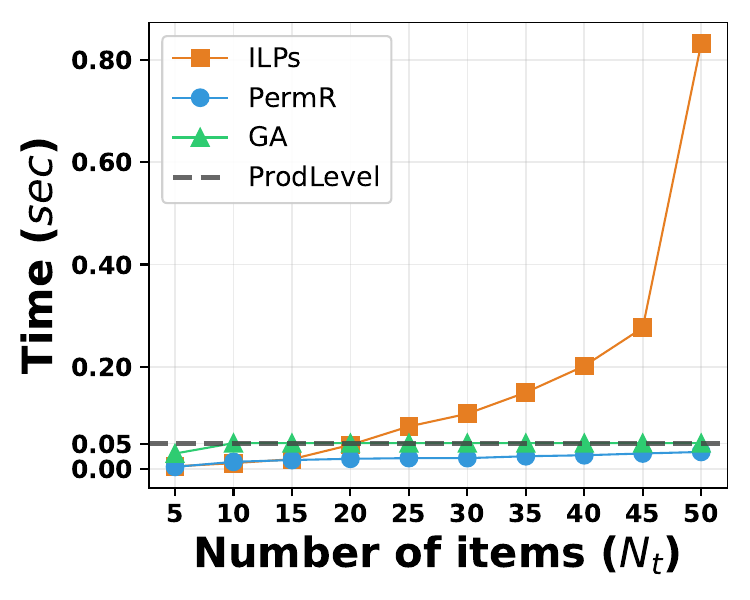}
         \caption{
         Mean inference time of ILPs, GA, PermR for different lengths of SERPs.
         }
         \label{fig:sub1}
     \end{subfigure}
     \hfill
     \begin{subfigure}[t]{0.48\linewidth}
         \centering
         \includegraphics[width=1.00\linewidth]{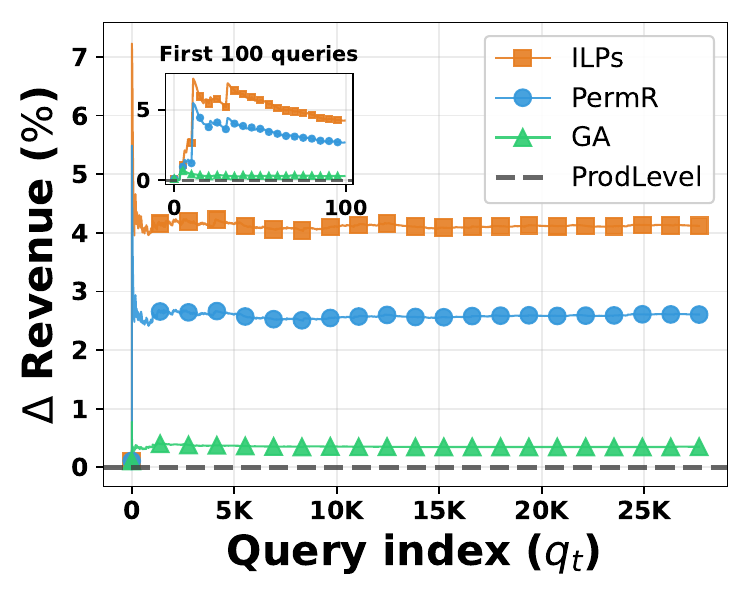}
         \caption{Moving average of revenue uplift under all queries. Inset zooms on the first 100 queries.}
         \label{fig:sub2}
     \end{subfigure}
     \caption{
     ILPs 
     achieve the maximum possible revenue uplift of $+4.1\%$, but exceed prod time limits by $0.05$ seconds for inference. 
     PermR 
     yields a $+2.6\%$ revenue uplift with no drop in time limit, while GA achieves only $+0.4\%$ revenue.
     }
     \label{fig:total_figure}
\end{figure}

Further, we assess the performance of ILPs (MOSEK on 12 threads), GA under 0.05 seconds, and PermR with $750$ iterations using a dataset of $27,000$ logged queries collected over 3 days in the production Goods category.
In our data, the majority of SERPs ($64\%$) have length $N = 50$. 
There are also samples with shorter lengths, enabling an additional evaluation of the inference time (as illustrated in Figure~\ref{fig:sub1}) of the ILPs and PermR for varying numbers of items: from 1 to 5 items, 6-10 items, \dots, up to 46-50 items.
The time complexity of PermR exhibits minimal growth with increasing SERP length, maintaining an acceptable production time of $0.05$ seconds.
If the solver's time complexity remains acceptable for SERPs with up to $N = 20$ items, it becomes unacceptable for the more typical cases of SERPs with $N = 50$ items.
In Figure \ref{fig:total_figure}, we evaluate the moving average of revenue uplift. We see that average revenue rapidly stabilizes at a horizontal plateau, indicating the robust stability of the solution.
Our results show that ILPs achieve the maximum possible revenue uplift of $+4.1\%$, GA only $+0.4\%$ while PermR yields a significant $+2.6\%$, which represents a $63\%$ approximation of the optimal solution.
In Table~\ref{tab::revenue_combined}, we provide additional insights into the revenue uplift of ILPs, GA, and  PermR for subcategories of Goods.
We ran PermR with $10$ different seeds, obtaining a consistent result of $2.607\% \pm 0.007\%$. Since we round to tenths, the seed can be fixed.
\subsection{Online Tests: PermR evaluation}
We deploy \textbf{PermR} with $I=750$ iterations and run an A/B test for 14 days.
The experiment covers 35\% of search traffic, corresponding to about 56 million queries. 
Online revenue is measured as the total platform pay-per-click products income, without position discount.
Table~\ref{tab::revenue_combined} shows that \textbf{PermR} increases revenue across most categories.
The largest gain is observed for the Home \& Garden subcategory (+6.0\%), while the Fashion subcategory shows the most stable improvement, with the narrowest confidence interval ($+4.2\pm 0.8\%$). 
Categories with higher offline potential tend to larger online gains. 
The ordering of mid-range categories varies, partly due to wider confidence intervals. 
PermR consistently increases revenue across categories while preserving all platform constraints.
\begin{table}[!h]
\centering
\caption{Revenue gain (\%) by category compared to the production ranking. 
\textbf{Offline}: $\Delta$ Revenue for ILP solvers, PermR, and genetic algorithm (GA) on logs. 
\textbf{Online}: $\Delta$ Revenue from a 14-day A/B test with PermR ($I{=}750$, 56M queries, 35\% traffic). Volume shows the query share in a category. The confidence intervals in A/B test are evaluated with p-value $<0.05$.}
\begin{tabular}{lcccccc}
\toprule
& \multicolumn{4}{c}{\textbf{Offline}} & \multicolumn{2}{c}{\textbf{Online A/B}} \\
\cmidrule(lr){2-5} \cmidrule(lr){6-7}
Subcategory & Vol.\ (\%) & ILPs & PermR & GA & Vol.\ (\%) & PermR \\
\midrule
All Goods             & 100 & $+4.1$ & $+2.6$ & $+0.4$ & 100 & $+2.0 \pm 0.4$ \\
\cmidrule(lr){2-5} \cmidrule(lr){6-7}
Fashion              & 25.3 & $+5.2$ & $+3.0$ & $+0.3$& 23.2 & $+4.2 \pm 0.8$ \\
SpareParts           & 17.7 & $+4.1$ & $+2.9$ & $+0.6$& 15.0 & $+2.4 \pm 0.7$ \\
Home \& Garden       &  7.1 & $+8.7$ & $+6.6$ & $+0.8$&  7.9 & $+6.0 \pm 1.8$ \\
AudioVideo           &  4.2 & $+3.5$ & $+2.0$ & $+0.2$&  5.9 & $+3.2 \pm 1.2$ \\
Info.\ Technology    &  4.3 & $+2.9$ & $+1.6$ & $+0.2$&  5.9 & $+1.9 \pm 1.3$ \\
Animals              &  4.2 & $+6.7$ & $+3.6$ & $+0.2$&  6.1 & $+4.8 \pm 1.5$ \\
Sports               &  3.3 & $+3.4$ & $+2.0$ & $+0.2$&  5.3 & $+3.7 \pm 0.9$ \\
Telecom              &  2.5 & $+4.2$ & $+2.5$ & $+0.2$& 10.6 & $+3.0 \pm 1.0$ \\
Hobby                &  5.2 & $+4.6$ & $+3.0$ & $+0.4$&  7.1 & $+2.6 \pm 1.3$ \\
Furniture            &  4.8 & $+0.7$ & $+0.4$ & $+0.1$&  5.3 & $+0.8 \pm 1.1$ \\
\bottomrule
\end{tabular}
\label{tab::revenue_combined}
\end{table}

\section{Conclusion}
This paper considers the reranking problem, aiming to increase the revenue of the e-commerce platform while preserving other metrics.
We formulate this problem as an Integer Linear Programming problem and study the efficiency of open-source solvers for the target problem dimensions.
There are two limitations: the infeasible long inference time of the solvers for deployment to the real-time ranking system, and the too low revenue uplift provided by the genetic algorithm from the open-source pymoo framework.
Therefore, we propose the PermR algorithm that approximates the solution while preserving the constraints.
Through extensive offline and online experiments, we demonstrate that PermR offers a reasonable trade-off between inference time and efficiency and is suitable for deployment in the product. 
A/B tests confirm that PermR increases the revenue by up to 2\% on average while preserving other metrics.

\section*{Acknowledgments}
This work was supported by The Ministry of Economic Development of the Russian Federation in accordance with the subsidy agreement (agreement identifier 000000C313925P4H0002; grant No 139-15-2025-012).

\bibliographystyle{unsrt}
\bibliography{sample-base}

\end{document}